\documentclass[10pt]{iopart}
\usepackage{amssymb, graphicx, cite, color, hyperref}

\newcommand{\F}{\mathcal{F}}
\renewcommand{\L}{\mathcal{L}}
\newcommand{\beq}{\begin{equation}}
\newcommand{\eeq}{\end{equation}}
\newcommand{\s}{\sigma}

\newcommand{\rmf}{\mathrm{f}}
\newcommand{\rmb}{\mathrm{b}}
\newcommand{\fub}{f_\mathrm{ub}}
\newcommand{\fuf}{f_\mathrm{uf}}
\newcommand{\lbar}{\bar{\ell}}
\newcommand{\lvar}{\ell_\mathrm{var}}
\newcommand{\lmin}{\ell_{\min}}
\newcommand{\Spath}{S_\mathrm{path}}
\newcommand{\nseq}{n_\mathrm{seq}}
\newcommand{\EOb}{E_\rmb^0}
\newcommand{\EObone}{E_{\rmb_1}^0}
\newcommand{\EObtwo}{E_{\rmb_2}^0}
\newcommand{\EOf}{E_\rmf^0}
\newcommand{\A}{\mathsf{A}}
\newcommand{\bigo}{\mathcal{O}}
\newcommand{\eps}{\epsilon}

\newcommand{\Ef}{E_\rmf}
\newcommand{\Eb}{E_\rmb}
\newcommand{\eff}{\mathrm{eff}}
\newcommand{\phys}{\mathrm{phys}}
\newcommand{\red}[1]{{#1}}

\begin{document}

\title[Scaling properties of evolutionary paths]{Scaling properties of evolutionary paths in a biophysical model of protein adaptation}

\author{Michael Manhart$^{1}$\footnote{Current address: Department of Chemistry and Chemical Biology, Harvard University, Cambridge, MA 02138, USA} and Alexandre V Morozov$^{1,2}$}

\address{$^1$ Department of Physics and Astronomy, Rutgers University, Piscataway, NJ 08854, USA} 
\address{$^2$ BioMaPS Institute for Quantitative Biology, Rutgers University, Piscataway, NJ 08854, USA}

\ead{morozov@physics.rutgers.edu}
\vspace{10pt}

\begin{abstract}
The enormous size and complexity of genotypic sequence space frequently requires consideration of coarse-grained sequences in empirical models.  We develop scaling relations to quantify the effect of this coarse-graining on properties of fitness landscapes and evolutionary paths.  We first consider evolution on a simple Mount Fuji fitness landscape, focusing on how the length and predictability of evolutionary paths scale with the coarse-grained sequence length and alphabet.  We obtain simple scaling relations for both the weak- and strong-selection limits, with a non-trivial crossover regime at intermediate selection strengths.  We apply these results to evolution on a biophysical fitness landscape that describes how proteins evolve new binding interactions while maintaining their folding stability.  We combine the scaling relations with numerical calculations for coarse-grained protein sequences to obtain quantitative properties of the model for realistic binding interfaces and a full amino acid alphabet.
\end{abstract}

%
%
%
%
\ioptwocol
%

\section{Introduction}


     The enormous size and dimensionality of genotypic sequence space are among the most salient features of molecular evolution.  These features not only present technical challenges for experiments and computation, but raise major conceptual questions as well: how can populations efficiently find high-fitness states in such a large space?  John Maynard Smith famously tackled this issue~\cite{MaynardSmith1970}, arguing that positive selection acting on individual mutations is key to efficiently evolving functional protein sequences.  However, this argument depends crucially on the structure of the fitness landscape and the underlying evolutionary dynamics.  One expects a large population to ascend a steep and perfectly-smooth landscape quickly, while substantial landscape ruggedness or genetic drift will slow down adaptation.
     
     The effect of ruggedness (due to epistatic interactions among genetic loci) on evolutionary paths has been a major focus of previous work.  These studies have investigated both simple models of fitness landscapes --- especially the uncorrelated random landscape~\cite{Kauffman1987, Flyvbjerg1992, Rokyta2006, Franke2011} (also known as the ``House of Cards''~\cite{Kingman1978}) and the rough Mount Fuji model~\cite{Aita2000, Franke2011, Neidhart2014} --- as well as landscapes empirically measured in specific organisms~\cite{Weinreich2006, Szendro2013a}.  
Populations in these studies are generally assumed to be under strong selection, so that evolutionary paths proceed strictly upward in fitness; a major goal is to determine the number and length of the accessible paths for different landscape topographies.  More recent work has begun to consider the effect of population dynamics (e.g., clonal interference) on evolutionary predictability~\cite{Szendro2013b}, a topic of central importance in evolutionary biology~\cite{Gould1990, Lobkovsky2012}.

     In most cases the computational and experimental cost of analyzing empirical models has required simplified sequence spaces, especially binary sequences (indicating only the presence or absence of a mutation at each site)~\cite{Flyvbjerg1992, Weinreich2006, Franke2011, Neidhart2014}, genomes or proteins with reduced lengths~\cite{Bloom2006, Neher2009, Lobkovsky2011}, and reduced alphabets of amino acids~\cite{Lobkovsky2011, Manhart2015} or protein structural components~\cite{Bogarad1999}.  However, it is not clear how properties of landscapes and evolutionary paths change under these implicit coarse-graining schemes.  Understanding their scaling behavior is essential for extending these models to more realistic biological systems.  Specifically, we must determine how properties of a model scale with both the coarse-grained sequence length $L$ and the coarse-grained alphabet size $k$ (number of possible alleles at each site), the latter being important when multiple mutations at a single site are likely.
     
     We first carry out this approach in a simple model of monomorphic populations undergoing substitutions on a smooth Mount Fuji landscape, showing how the scaling properties of the model depend crucially on the strength of selection relative to genetic drift.  We then consider evolution on a fitness landscape based on the biophysics of protein folding and binding, describing how proteins evolve new binding interactions while maintaining folding stability~\cite{Manhart2015}.  Using scaling relations, we are able to extend numerical calculations carried out for coarse-grained representations of proteins, obtaining quantitative evolutionary properties for realistic binding interface sizes and a full amino acid alphabet.

\section{Evolutionary paths on a smooth Mount Fuji landscape}

     We first consider a simple fitness landscape model, the smooth ``Mount Fuji'' (i.e., single-peaked) landscape~\cite{Kauffman1989}.  Consider genotypic sequences of length $L$ with $k$ possible alleles $\{\A_1, \A_2, \ldots, \A_k\}$ at each site, resulting in $\nseq = k^L$ possible genotypes.  We assume the alleles $\{\A_1, \A_2, \ldots, \A_k\}$ are in increasing order of fitness rank.  The sites could be residues in a protein, nucleotides in a DNA sequence, or larger genomic loci such as whole genes.  In general we will interpret the sequences in the model as coarse-grained versions of actual biological sequences.  For example, a 12-residue binding interface on a protein with 20 possible amino acids at each site could be coarse-grained into $L=6$ pairs of sites with $k=5$ alleles at each site, where each allele represents a class of amino acids grouped by physico-chemical properties (e.g., negative, positive, polar, hydrophobic, and other).  This is analogous to block spin renormalization in Ising models~\cite{Yeomans1992}.
     
     Let the occupation number $n_j(\s)$ of a sequence $\s$ be the number of $\A_j$ alleles in the sequence, so that $\sum_{j=1}^k n_j(\s) = L$.  We define the fitness of a sequence $\s$ to be
     
\beq
\F(\s) = f^{\sum_{j=1}^k (j-1) n_j(\s)}, 
\label{eq:MF_fitness}
\eeq

\noindent where $f \ge 1$ is the minimum multiplicative fitness change from a single mutation: a mutation $\A_i \to \A_j$ at a single site changes fitness by a factor of $f^{j-i}$.  If $f = 1$, the fitness landscape is flat and evolution is neutral, while if $f > 1$, the landscape has a minimum at $\s = \A_1 \A_1 \cdots \A_1$ ($\F = 1$) and a maximum at $\s = \A_k \A_k \cdots \A_k$ ($\F = f^{L(k-1)}$).  The model is non-epistatic since the fitness function factorizes over sites; thus all mutations have the same fitness effect regardless of the genetic background on which they occur.  A more general Mount Fuji model could allow mutations at different sites and between different alleles to have different fitness effects, although this will not affect the scaling properties of the model that are of primary interest here.

     We assume that the population is monomorphic: all organisms have the same genotype at any given time.  This approximation holds when $u \ll (LN\log N)^{-1}$, where $u$ is the per-site probability of mutation per generation and $N$ is the population size~\cite{Champagnat2006}.  In this regime the population evolves through a series of substitutions, in which single mutants arise and fix one at a time.  A substitution from genotype $\s$ to $\s'$ occurs at the rate~\cite{Kimura1983}
     
\beq
W(\s'|\s) = Nu~\phi(s),
\label{eq:sub_rate}
\eeq

\noindent where $\phi(s)$ is the fixation probability of a single mutant with selection coefficient $s = \F(\s')/\F(\s) - 1$.  We use the diffusion approximation to the Wright-Fisher model for the fixation probability~\cite{Kimura1962}: 

\beq
\phi(s) = \frac{1 - e^{-2s}}{1 - e^{-2Ns}}.
\label{eq:fix_prob}
\eeq

\noindent Note that when $N|s| > 1$ this can be approximated by

\beq
\phi(s) \approx \left\{
\begin{array}{ll}
1 - e^{-2s} & \mathrm{if} \, s > 0, \\
0 & \mathrm{if} \, s < 0. \\
\end{array}
\right.
\label{eq:fix_prob_inf_N}
\eeq

\noindent That is, when selection is much stronger than genetic drift, deleterious mutations never fix, while beneficial mutations fix with a probability commensurate with their selective advantage.  This is often referred to as the ``strong-selection weak-mutation'' (SSWM) limit~\cite{Gillespie1984}.


\subsection{The ensemble of evolutionary paths}

     For concreteness we consider the following evolutionary process: the population begins at the least fit genotype, $\A_1 \A_1 \cdots \A_1$, and evolves according to~\eref{eq:sub_rate} until it reaches the most fit genotype, $\A_k \A_k \cdots \A_k$, for the first time.  Define an evolutionary path $\varphi$ as the ordered sequence of genotypes $\varphi = (\s_0, \s_1, \ldots, \s_\ell)$ traversed by the population during this process, where $\s_0 = \A_1 \A_1 \cdots \A_1$ and $\s_\ell = \A_k \A_k \cdots \A_k$.  The probability of making a single substitution $\s \to \s'$, given that a substitution out of $\s$ occurs, is

\beq
Q(\s'|\s) = W(\s'|\s) ~ \theta(\s),
\label{eq:jump_prob}
\eeq

\noindent where $\theta(\s) = \left( \sum_{\s'} W(\s'|\s) \right)^{-1}$ is the mean waiting time in $\s$ before a substitution occurs.  Thus the probability of taking a path $\varphi$ is
     
\beq
\Pi[\varphi] = \prod_{i=0}^{\ell-1} Q(\s_{i+1}|\s_i).
\eeq

\noindent Since the population is guaranteed to reach the final state eventually, $\sum_\varphi \Pi[\varphi] = 1$, where the sum is over all first-passage paths $\varphi$ between the initial and final states.

     We are interested in statistical properties of this evolutionary path ensemble.  We can calculate many such properties using an exact numerical algorithm described in \ref{sec:algorithm}~\cite{Manhart2013, Manhart2014}.  Here we are especially interested in the distribution of path lengths $\ell$, i.e., the number of substitutions experienced by the population before it first reaches the fitness maximum. The path length distribution $\rho(\ell)$ is defined as

\beq
\rho(\ell) = \sum_\varphi \delta_{\ell,\L[\varphi]} ~ \Pi[\varphi],
\eeq

\noindent where $\L[\varphi]$ is the length of path $\varphi$, and $\delta$ is the Kronecker delta.  We can similarly express the mean $\lbar$ and variance $\lvar$ of path length.  We also consider the path entropy $\Spath$, defined as
     
\beq
\Spath = - \sum_\varphi \Pi[\varphi] \log \Pi[\varphi].
\eeq

\noindent This quantity measures the predictability of evolution in sequence space: if only a single path is accessible, $\Spath = 0$, and evolution is perfectly predictable.  Larger values of $\Spath$, on the other hand, indicate a more diverse ensemble of accessible pathways, and thus less predictable evolution.


\subsection{Neutral limit}

     We first consider properties of the evolutionary path ensemble in the case of neutral evolution ($f = 1$ in~\eref{eq:MF_fitness}).  For simple random walks on finite discrete spaces, previous work has shown that the mean first-passage path length scales with the total number of states~\cite{Bollt2005, Condamin2007}, while the distribution of path lengths is approximately exponential~\cite{Bollt2005}.  Thus for neutral evolution, 
     
\beq
\lbar \sim \nseq = k^L, \qquad \lvar \sim \lbar^{~2} \sim k^{2L}.
\label{eq:neutral_lbar}
\eeq

\noindent 
Conceptually, this means the population on average must explore the entire sequence space before reaching a particular point for the first time, and thus the average number of substitutions grows exponentially with the length of the sequence.  Moreover, since the standard deviation is of the same order as the mean, paths much longer than the mean are likely.

     Let $\gamma$ be the average connectivity, defined as the average number of single-mutant substitutions accessible from each sequence; in neutral evolution all single-mutant substitutions are accessible, so $\gamma = L(k-1)$.  Since all substitutions are equally likely, $Q(\s'|\s) = \gamma^{-1}$ for $\s$ and $\s'$ separated by a single mutation.  The entropy of the neutral path ensemble is therefore~\cite{Manhart2014}

\begin{eqnarray}
\Spath & = - \sum_\varphi \Pi[\varphi] \log \gamma^{-\L[\varphi]}, \nonumber\\
& = \lbar \log \gamma, \nonumber\\
& \sim k^L \log L(k-1).
\label{eq:neutral_Spath}
\end{eqnarray}

\noindent The path entropy consists of two distinct components: the average path length and the average connectivity.  The factor of $\log\gamma$ is the average entropy contribution from each jump in the path.
It is worth noting that mean path length (and the distribution of path lengths in general) does \emph{not} have explicit dependence on connectivity: it only depends on the size of the space.  So it is the enormous size, not the connectivity, of sequence space that causes neutral evolution to require so many steps to reach a particular point.  In contrast, path entropy, and thus evolutionary predictability, depends on \emph{both} the size and connectivity of sequence space.


\subsection{Strong-selection limit}

     We now consider evolutionary paths in the strong-selection limit.  Here all beneficial mutations are selected so strongly ($f \gg 1$ in~\eref{eq:MF_fitness}) that their fixation probabilities are all approximately 1, while deleterious mutations never occur.  Thus evolutionary paths proceed strictly upward on the fitness landscape.  \red{This is sometimes called the ``adaptive walk''~\cite{Kauffman1987} or ``random adaptation''~\cite{Orr2002} scenario;} it is identical to zero-temperature Metropolis Monte Carlo dynamics with energy replaced by negative fitness~\cite{Flyvbjerg1992}.  
Since the fitness landscape is non-epistatic and reverse mutations are impossible, each site can be considered to evolve independently.  In particular, we can decompose the total path length into a sum of path lengths for individual sites, so that the path length cumulants for the entire sequence are simply sums of the cumulants for individual sites.  (Note that the restriction to first-passage paths effectively couples all the sites because they must all reach their final states simultaneously, and so site independence is only valid when reverse mutations are prohibited.)

     In \ref{sec:lbar_derivation} we show that \red{the mean number of substitutions for a single site in this limit is $\lbar = H_{k-1}$ (the $(k-1)$th harmonic number), consistent with previous results~\cite{Gillespie1983, Orr2002}.}  Hence the mean length for $L$ sites is $LH_{k-1}$, and since $H_{k-1} = \log k + b + \bigo(k^{-1})$, the mean length scales as
     
\beq
\lbar \sim L(\log k + b).
\label{eq:ss_lbar}
\eeq

\noindent We explicitly include the $\bigo(1)$ constant $b$ here since it may be comparable to $\log k$ if $k$ is not too large.  For the harmonic numbers, $b$ is equal to the Euler-Mascheroni constant $\gamma_\mathrm{EM} \approx 0.5772$, but we use generic notation here since we will fit this same scaling form to an empirical model in the next section.  
\Eref{eq:ss_lbar} implies that $\lbar$ scales approximately logarithmically with the size $\nseq$ of sequence space, compared with the linear scaling seen in the neutral case~\eref{eq:neutral_lbar}.  Moreover, \ref{sec:rho_derivation} shows that $\rho(\ell)$ is approximately Poisson, and thus the variance $\lvar$ should obey the same scaling as $\lbar$.

     The average connectivity of sequence space is reduced compared to the neutral case, since only beneficial substitutions are allowed.  The connectivity averaged over all sequences is $L(k-1)/2$ (\ref{sec:size_connectivity}); the reduction by a factor of 2 is intuitively explained by the fact that every allowed beneficial substitution has a prohibited deleterious substitution.  For the path entropy under strong selection, we take as an ansatz the same dependence on $\lbar$ and $\gamma$ as in~\eref{eq:neutral_Spath}, albeit with different $L$, $k$ scaling:

\beq
\Spath \sim \lbar \log \gamma \sim L(\log k + b) \log \frac{1}{2} L(k-1).
\eeq

\noindent We numerically verify this ansatz in the next section (\fref{fig:simple_model}).


\subsection{Coarse-graining and landscape-dependence of scaling relations}

     The path scaling relations depend qualitatively on whether the fitness landscape is flat (neutral evolution) or very steep (strong selection).  How does the transition between these two limits occur at intermediate selection strengths, where selection and stochastic fluctuations (genetic drift) compete more equally?  We now implement a renormalization scheme for coarse-graining sequence space on the fitness landscape of~\eref{eq:MF_fitness}.  Let $s$ be the total selection coefficient between the minimum and maximum fitness points on the landscape; \red{this corresponds to the actual selection strength between two distinct biological genotypes.  For example, the minimum and maximum fitnesses might correspond to wild-type and antibiotic-resistant genotypes in bacteria~\cite{Weinreich2006, Schenk2013}, or to one protein sequence that does not bind a target ligand and one that does~\cite{Manhart2015}.  As we coarse-grain the sequence space into smaller $L$ and $k$, we must therefore hold fixed this true overall selection strength.  Since $s = f^{L(k-1)} - 1$ in the Mount Fuji model~\eref{eq:MF_fitness}, we renormalize the minimum fitness benefit $f$ accordingly:}
     
\beq
f = (1 + s)^{1/(L(k-1))}.
\eeq

\noindent Thus the fitness benefit of each mutation increases \red{as we coarse-grain the sequence space (decrease $L$ and $k$), since each mutation in the model corresponds to several mutations on the true biological sequences.}
     
\begin{figure*}
\centering\includegraphics[width=\textwidth]{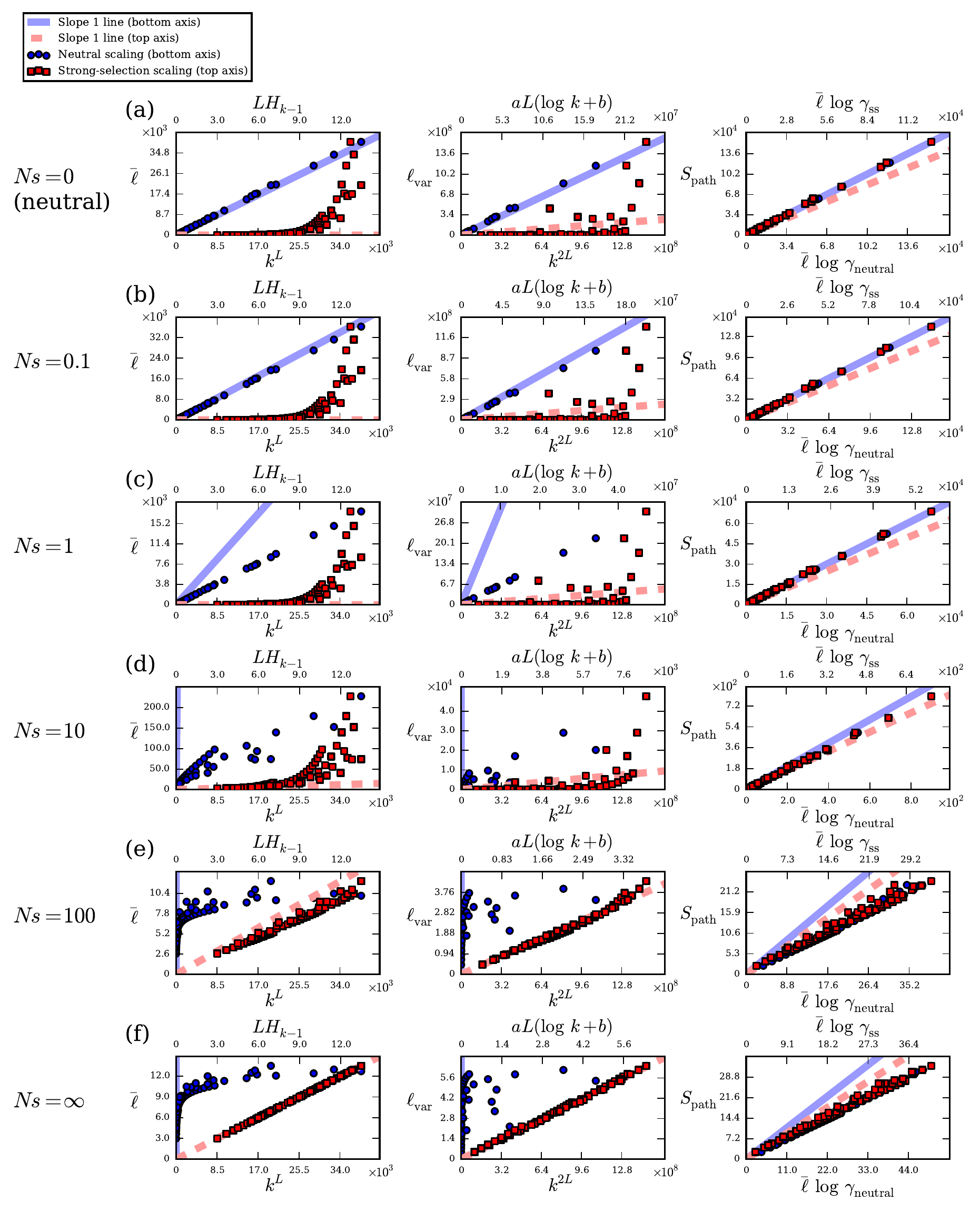}
\caption{
Scaling properties of evolutionary paths on the Mount Fuji landscape \eref{eq:MF_fitness} for different values of $Ns$, where $N = 1000$ and $s$ is the total selection coefficient from the least fit to the most fit sequence on the landscape: 
(a)~$Ns = 0$ (neutral evolution), 
(b)~$Ns = 0.1$, 
(c)~$Ns = 1$, 
(d)~$Ns = 10$, 
(e)~$Ns = 100$, and 
(f)~$Ns = \infty$. 
The left column shows mean path length $\lbar$ (number of substitutions), the middle column shows path length variance $\lvar$, and the right column shows path entropy $\Spath$.  Each panel plots numerical data against both neutral scaling parameters on the bottom axes (blue circles and the solid blue line of slope 1; $\gamma_\mathrm{neutral} = L(k-1)$), as well as strong-selection scaling parameters on the top axes (red squares and the dashed red line of slope 1; $\gamma_\mathrm{ss} = L(k-1)/2$).  Numerical values of the variance $\lvar$ are fitted to a function of the form $aL(\log k + b)$ for each value of $Ns$ separately.  We scan over all $L > 1$ and $k > 2$ such that $k^L < 4 \times 10^4$. 
}
\label{fig:simple_model}
\end{figure*}

     We consider a range of total $s$ values and numerically calculate path statistics for each $L$ and $k$ using the method of \ref{sec:algorithm}.  In \fref{fig:simple_model} we show the scaling of $\lbar$, $\lvar$, and $\Spath$ calculated in this manner for several values of relative selection strength $Ns$.  For $Ns = 0$, we not only confirm the neutral scaling relations~\eref{eq:neutral_lbar} but also observe that any proportionality factors and additive constants are so negligible that the scaling relations are actually approximate equalities (\fref{fig:simple_model}a).  The predicted relation for the path entropy \eref{eq:neutral_Spath} also holds exactly.  Moreover, weak selection appears to preserve these scaling relations: they still hold even at $Ns = 0.1$ (\fref{fig:simple_model}b).  When selection becomes comparable to genetic drift ($Ns = 1$, \fref{fig:simple_model}c), the neutral scaling relations still hold qualitatively, although the slopes of $\lbar \sim k^L$ and $\lvar \sim k^{2L}$ are no longer close to 1, indicating different proportionality factors.  
    
     At the other extreme ($Ns = \infty$,~\fref{fig:simple_model}f), the strong-selection scaling relations~\eref{eq:ss_lbar} for path length hold as expected.  We also verify that $\Spath \sim \lbar\log\gamma$ even for strong selection, albeit with a proportionality factor less than 1.  This scaling maintains at finite but large selection strengths of $Ns = 100$ (\fref{fig:simple_model}e).  At intermediate selection strengths ($Ns = 10$,~\fref{fig:simple_model}d), however, neither set of scaling relations for $\lbar$ and $\lvar$ holds, indicating that path length statistics are no longer simple functions of sequence space size $k^L$.

\section{Evolutionary paths in a biophysical model of protein adaptation}


     Simple model landscapes defined in genotype space, such as~\eref{eq:MF_fitness}, have produced many theoretical results and guided analysis of some data~\cite{Kauffman1987, Flyvbjerg1992, Rokyta2006, Franke2011, Szendro2013a, Neidhart2014}.  However, their purely phenomenological nature allows for little interpretation of their parameters and includes no basis in the underlying molecular processes --- interactions among proteins, DNA, RNA, and other biomolecules --- that govern cells.  Thus a promising alternative is to develop models of fitness that explicitly account for these molecular properties~\cite{Bloom2006, Lobkovsky2011, Schenk2013, Haldane2014, Serohijos2014b}.  We now consider the scaling properties of evolutionary paths in such a model based on the biophysics of protein folding and binding~\cite{Manhart2013, Manhart2014, Manhart2015}.  


\subsection{Protein energetics and coarse-graining}
\label{sec:protein_coarse_graining}

     Consider a protein with two-state folding kinetics~\cite{Creighton1992}.  In the folded state, the protein has an interface that binds a target molecule.  Because the protein can bind \emph{only} when it is folded, the protein has three possible structural states: folded and bound, folded and unbound, and unfolded and unbound.  Let the free energy of folding be $\Ef$ (often denoted by $\Delta G$), so that an intrinsically-stable protein has $\Ef < 0$.  Let the free energy of binding, relative to the chemical potential of the target molecule, be $\Eb$, so that $\Eb < 0$ indicates a favorable binding interaction. Note that $\Eb$ becomes more
favorable as the chemical potential of the target molecule is increased.

     The folding and binding energies depend on the protein's genotype (amino acid sequence) $\s$.  \red{We assume adaptation only affects the $L$ residues at the binding interface, which, to a first approximation, make additive contributions to the total folding and binding free energies~\cite{Wells1990}:}

\begin{eqnarray}
\Ef(\s) & = \EOf + \sum_{i = 1}^L \eps_\rmf(i, \s^i), \nonumber\\
\Eb(\s) & = \EOb + \sum_{i = 1}^L \eps_\rmb(i, \s^i),
\label{eq:Ef_Eb}
\end{eqnarray}

\noindent where $\eps_\rmf(i, \s^i)$ and $\eps_\rmb(i, \s^i)$ are entries of energy matrices that capture the energetic contributions of amino acid $\s^i$ at position $i$.  Folding and binding energetics are probed experimentally and computationally by measuring the changes (often denoted by $\Delta\Delta G$) in $\Ef$ or $\Eb$ resulting from single-point mutations~\cite{Bogan1998, Thorn2001, Tokuriki2007}.  \red{These studies generally indicate that each position makes an energetic contribution of order $1$ kcal/mol to the total energy.  As a simple approximation, we sample each energy contribution $\eps_{\rmf,\rmb}(i, \s^i)$ independently from a Gaussian distribution with zero mean and standard deviation 1 kcal/mol.  The offsets $\EOf$ and $\EOb$ therefore correspond to the average folding and binding energies of the protein with a random sequence at the binding interface; $\EOf$ includes the folding stability contribution from all residues in the protein away from the binding interface.  As long as it produces a physically realistic range of total energies, the exact shape of the distributions for $\eps_{\rmf,\rmb}(i, \s^i)$ is unimportant for large enough $L$ due to the central limit theorem.}

     
     \red{Numerical calculations over all $k^L$ sequences are not possible for large $L$ and a full amino acid alphabet ($k=20$).  However, we can consider coarse-grained versions of the model by grouping positions and amino acids into classes, resulting in some effective sequence parameters $L_\eff$ and $k_\eff$ that are smaller than their physical counterparts $L_\phys$ and $k_\phys = 20$.  If we then determine how properties of the model scale with $L_\eff$ and $k_\eff$ under such a coarse-graining procedure, we can extrapolate these properties to the physical values $L_\phys$ and $k_\phys$.
     
     As we vary $L_\eff$ and $k_\eff$, we must renormalize the distributions of energetic contributions $\eps_{\rmf,\rmb}(i, \s^i)$ for the effective sequences such that the distribution of total sequence energies remains constant, similar to our coarse-graining scheme in the previous section.  Since the total sequence energies are sums of Gaussian contributions from each site~\eref{eq:Ef_Eb}, coarse-graining the sites amounts to sampling the effective $\eps_{\rmf,\rmb}(i, \s^i)$ values from a Gaussian distribution with standard deviation rescaled by a factor of $\sqrt{L_\phys/L_\eff}$.  For example, if $L_\phys = 12$ and $L_\eff = 6$ (grouping positions into pairs), then each effective $\eps_{\rmf,\rmb}(i, \s^i)$ is the sum of two physical $\eps_{\rmf,\rmb}(i, \s^i)$ values, and hence the effective $\eps_{\rmf,\rmb}(i, \s^i)$ should have zero mean and standard deviation $\sqrt{2}$ kcal/mol.  Note that we analytically continue this rescaling to consider values of $L_\eff$ and $k_\eff$ that do not evenly divide $L_\phys$ and $k_\phys$.}  For simplicity we will drop the ``eff'' labels and hereafter interpret $L$, $k$, $\eps_\rmf(i, \s^i)$ and $\eps_\rmb(i, \s^i)$ as these effective, coarse-grained parameters unless indicated otherwise.

     


\subsection{Evolutionary model}

     Without loss of generality, we assume the protein contributes fitness 1 to the organism when it is both folded and bound.  Let $\fub, \fuf \in [0, 1]$ be the multiplicative fitness penalties for being unbound and unfolded, respectively: the fitness is $\fub$ if the protein is unbound but folded, and $\fub\fuf$ if the protein is both unbound and unfolded.  Then the fitness of the protein averaged over all three possible structural states is given by~\cite{Manhart2015}

\beq
\F(\Ef, \Eb) = \frac{e^{-\beta(\Ef + \Eb)} + \fub e^{-\beta \Ef} + \fub\fuf}{e^{-\beta(\Ef + \Eb)} + e^{-\beta \Ef} + 1},
\label{eq:protein_fitness}
\eeq

\noindent where $\beta = 1.7 ~(\mathrm{kcal/mol})^{-1}$ is inverse room temperature and the structural states are assumed to be in thermodynamic equilibrium.

     We assume that the population begins as perfectly adapted to binding a target molecule characterized by energy matrix $\eps_{\rmb_1}$ with offset $\EObone$ (defining a fitness landscape $\F_1$).  The population is then subjected to a selection pressure which favors binding a new target, with energy matrix $\eps_{\rmb_2}$ and offset $\EObtwo$ (fitness landscape $\F_2$).  \red{We assume that the binding energy matrices for the new and old targets are uncorrelated, although this assumption is not essential.}  The population evolves in the monomorphic limit with the SSWM dynamics in~\eref{eq:sub_rate} and~\eref{eq:fix_prob_inf_N}.  Thus the evolutionary paths are first-passage paths leading from the genotype corresponding to the global maximum on $\F_1$ to a local or global maximum on $\mathcal{F}_2$, with fitness increasing monotonically along each path.


\subsection{Case 1: selection for binding strength}

\begin{figure*}[tbph!]
\centering\includegraphics[width=0.8\textwidth]{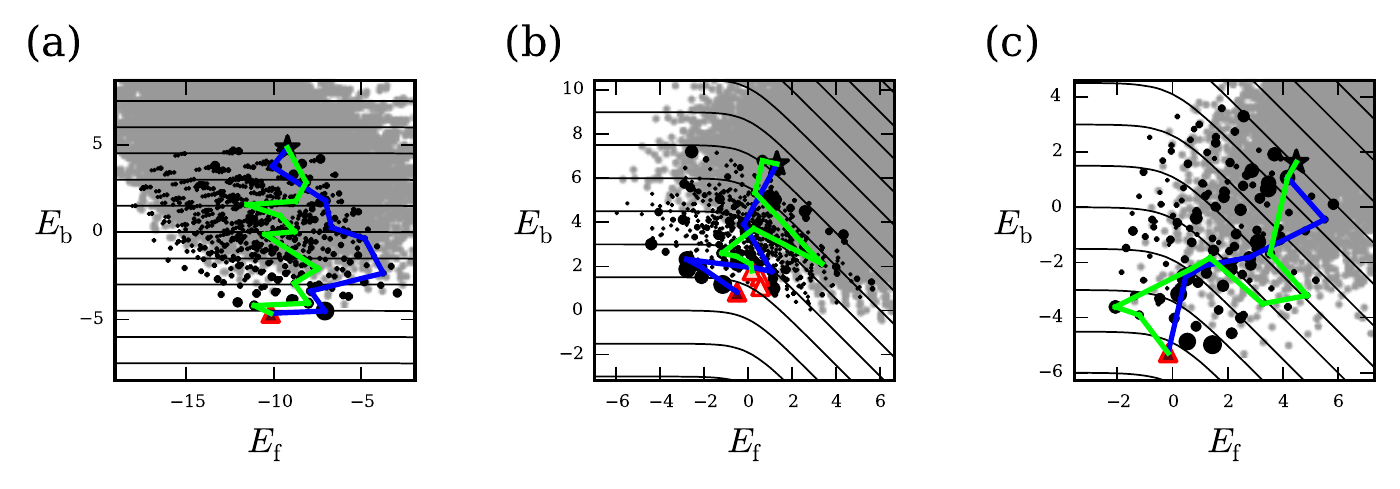}
\caption{
Example landscapes of protein adaptation with direct selection for binding only (case 1), zoomed into the region of energy space accessible to evolutionary paths in our model.  
(a)~Stable protein with $\EOf = -10$ kcal/mol, 
(b)~marginally-stable protein with $\EOf = 3$ kcal/mol, and
(c)~intrinsically-unstable protein with $\EOf = 8$ kcal/mol. 
In all panels $\fub = 0$, $\fuf = 1$, and $\EObone = \EObtwo = 5$ kcal/mol.  The coarse-grained sequence parameters are $L = 6$ and $k=5$, with \red{effective energy matrices $\eps_\rmf$, $\eps_{\rmb_1}$, and $\eps_{\rmb_2}$ sampled from distributions that were rescaled using $L_\phys = 12$.}
The black star indicates the initial state for adaptation (global maximum on $\F_1$); red triangles indicate local fitness maxima on $\F_2$, shaded according to their commitment probabilities (probability of reaching that final state starting from the initial state); black circles indicate intermediate states along paths, sized proportional to their path density (total probability of paths passing through them); small gray circles are genotypes inaccessible to adaptation.  The black contours indicate constant fitness $\F_2$ (the fitness difference between adjacent contours is non-uniform so that they are equidistant in energy space), while example paths are shown in blue and green.
}
\label{fig:case1_landscapes}
\end{figure*}

     There are three qualitatively distinct cases of the fitness landscape in~\eref{eq:protein_fitness}, depending on the values of the parameters $\fub$ and $\fuf$~\cite{Manhart2015}.  These cases correspond to different biological scenarios for the selection pressures on binding and folding.  In the simplest scenario (``case 1''), proteins are selected for their binding function ($\fub < 1$), but misfolding carries no additional fitness penalty (e.g., due to toxicity of misfolded proteins) beyond loss of function ($\fuf = 1$).  Thus we say there is direct selection for binding only.  Three examples of adaptation in this regime are shown in \fref{fig:case1_landscapes}; the main determinant of the qualitative nature of adaptation is the overall folding stability $\Ef$.
     
     Although the model is non-epistatic at the level of the energy traits (since \eref{eq:Ef_Eb} is additive), there is epistasis at the level of fitness~\eref{eq:protein_fitness} due to its nonlinear dependence on energy.  Indeed, there is widespread magnitude epistasis, which occurs when the fitness effect of a mutation has different magnitude on different genetic backgrounds, although it is always beneficial or always deleterious.  Sign epistasis, which occurs when a mutation can be beneficial on one background but deleterious on another, manifests itself as curvature in the fitness contours in energy space~\cite{Manhart2015}, as shown in \fref{fig:case1_landscapes}.  However, we see that the landscape is largely free of sign epistasis except near $\Ef = 0$, where there is a higher probability of multiple local fitness maxima (\fref{fig:case1_landscapes}b).  Overall, this suggests that the scaling relations from the non-epistatic Mount Fuji model may provide a reasonable approximation for this model of protein adaptation; the approximately additive nature of protein traits as in~\eref{eq:Ef_Eb} has led to applications of the Mount Fuji model to proteins previously~\cite{Aita1996, Aita1998, Aita2000, Schenk2013}.

\begin{figure}[tbph!]
\centering\includegraphics[width=\columnwidth]{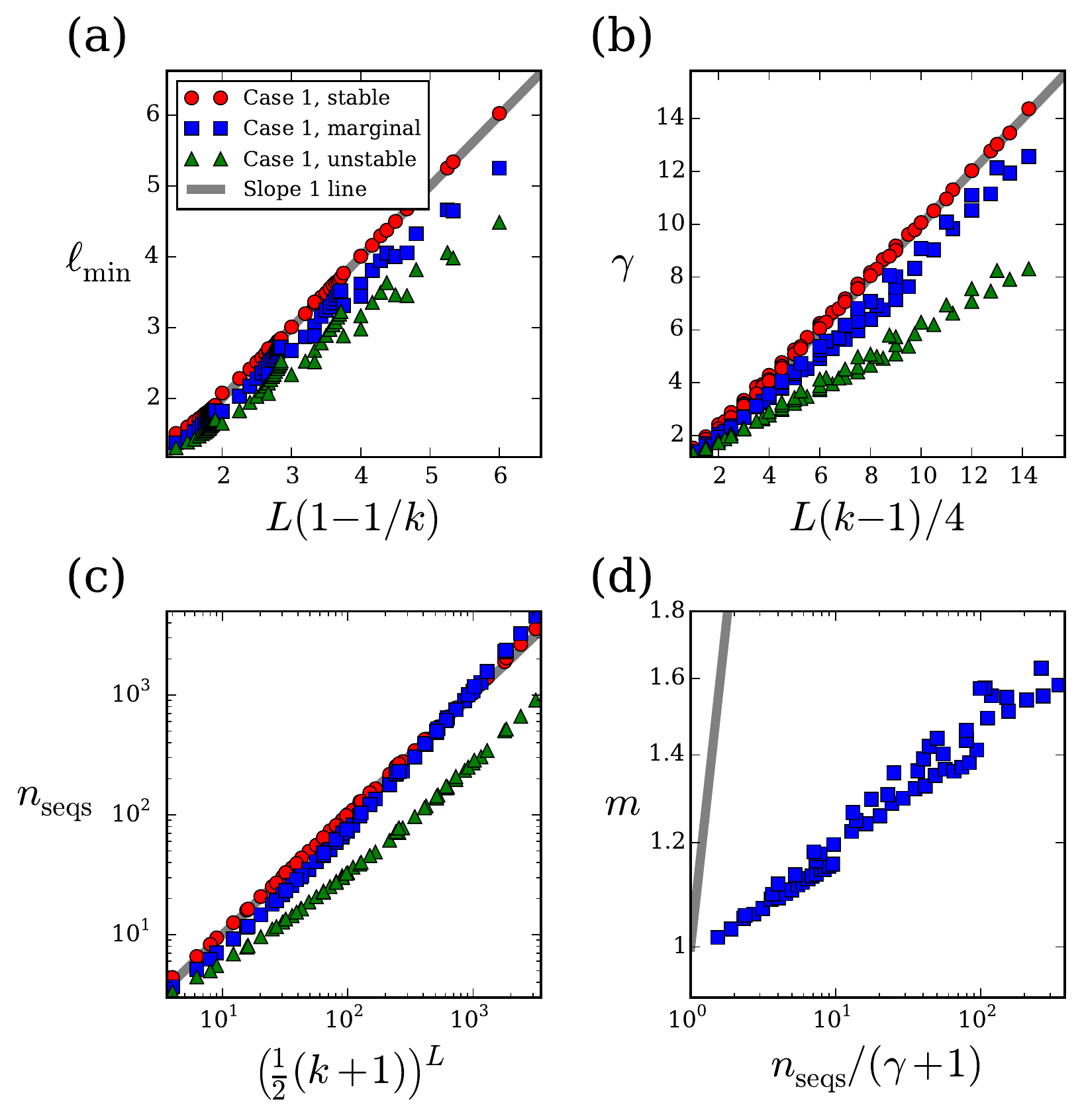}
\caption{
Scaling of landscape properties for three regimes of protein adaptation with direct selection for binding only (case 1). 
(a)~Minimum path length $\lmin$, equal to the Hamming distance between the initial and final states, versus $L(1-1/k)$;  
(b)~average connectivity $\gamma$ versus $L(k-1)/4$; 
(c)~average number $\nseq$ of accessible sequences versus $((k+1)/2)^L$; and 
(d)~average number $m$ of local fitness maxima versus $\nseq/(\gamma + 1)$.  
In all panels red circles are for stable proteins, blue squares are for marginally-stable proteins, and green triangles are for intrinsically-unstable proteins, with all energy and fitness parameters the same as in \fref{fig:case1_landscapes}.  Each point represents an average over $10^4$ realizations of the folding and binding energy matrices; we exclude trivial realizations where the initial state is already a local maximum on $\F_2$.  We scan over all $L > 1$ and $k > 2$ such that $k^L < 4 \times 10^4$, \red{with energy matrices rescaled using $L_\phys = 12$.}  
Slope 1 lines from the origin are shown in gray to guide the eye.  
}
\label{fig:case1_landscape_scaling}
\end{figure}

     In \fref{fig:case1_landscape_scaling} we show scaling properties of the genotypic fitness landscape for the three $\Ef$ regimes of the model for case 1 (corresponding to the examples in \fref{fig:case1_landscapes}).  The minimum path length $\lmin$ is the Hamming distance between the initial and final states for adaptation; for a randomly-chosen initial sequence, $\lmin = L(1 - 1/k)$ on average.  Indeed, this relation accurately describes the regime of stable proteins (\fref{fig:case1_landscape_scaling}a).  For proteins that are already sufficiently stable, there is no selection pressure to improve stability further, so the global fitness maximum is almost always the best-binding sequence.  Since the binding energetics for the old and new targets are uncorrelated, the initial and final states are therefore uncorrelated as well, which explains the $\lmin$ scaling.  For marginally-stable and unstable proteins, $\lmin$ still scales with $L(1 - 1/k)$, but with a reduced slope.  This is because the initial and final states become correlated in these two cases.  We can think of this effect as a reduction in the effective length $L$, since more beneficial mutations are already present in the initial state.  We see similar behavior in the average connectivity $\gamma$ and accessible size $\nseq$ of sequence space (\fref{fig:case1_landscape_scaling}b,c).  Note that a random initial state reduces the average connectivity of the accessible sequence space by an additional factor of 2, yielding $\gamma = L(k-1)/4$ (see \ref{sec:size_connectivity}).
     
     Whereas stable and unstable proteins almost always have a single fitness maximum, marginally-stable proteins have a sizable probability of multiple maxima owing to greater sign epistasis (\fref{fig:case1_landscapes}b). In a purely random, uncorrelated fitness landscape, the average number of local maxima is $m = k^L/(L(k-1) + 1)$~\cite{Kauffman1987}.  This has the form $\nseq/(\gamma + 1)$: the number of maxima increases with the total size of the space and decreases with the connectivity.  We empirically test this scaling for the average number of maxima in a marginally-stable protein, and we find good agreement (\fref{fig:case1_landscape_scaling}d).  By fitting numerically-calculated values of $m$ as a power law of $\nseq/(\gamma + 1)$, we obtain an anomalous scaling exponent of $\approx 0.09$; the fact this is much less than 1 reflects the highly-correlated nature of our fitness landscape.  The fitted scaling relation allows us to accurately determine the average number of local maxima for binding interfaces and amino acid alphabets much larger than we can directly calculate.  By also fitting $\gamma$ as a linear function of $L(k-1)/4$ (\fref{fig:case1_landscape_scaling}b) and $\nseq$ as a power law of $((k+1)/2)^L$ (\ref{sec:size_connectivity}, \fref{fig:case1_landscape_scaling}c), we estimate the number of local maxima to be $\approx 11$ for a marginally-stable protein with $L_\phys = 12$ binding interface residues and an amino acid alphabet of size $k_\phys = 20$.  \red{This number of maxima is much smaller than the total number of sequences ($k^L \approx 4 \times 10^{15}$) and the expected number of maxima on an uncorrelated random landscape of the same size ($k^L/(L(k-1) + 1) \approx 1.8 \times 10^{13}$).}
     
\begin{figure}
\centering\includegraphics[width=\columnwidth]{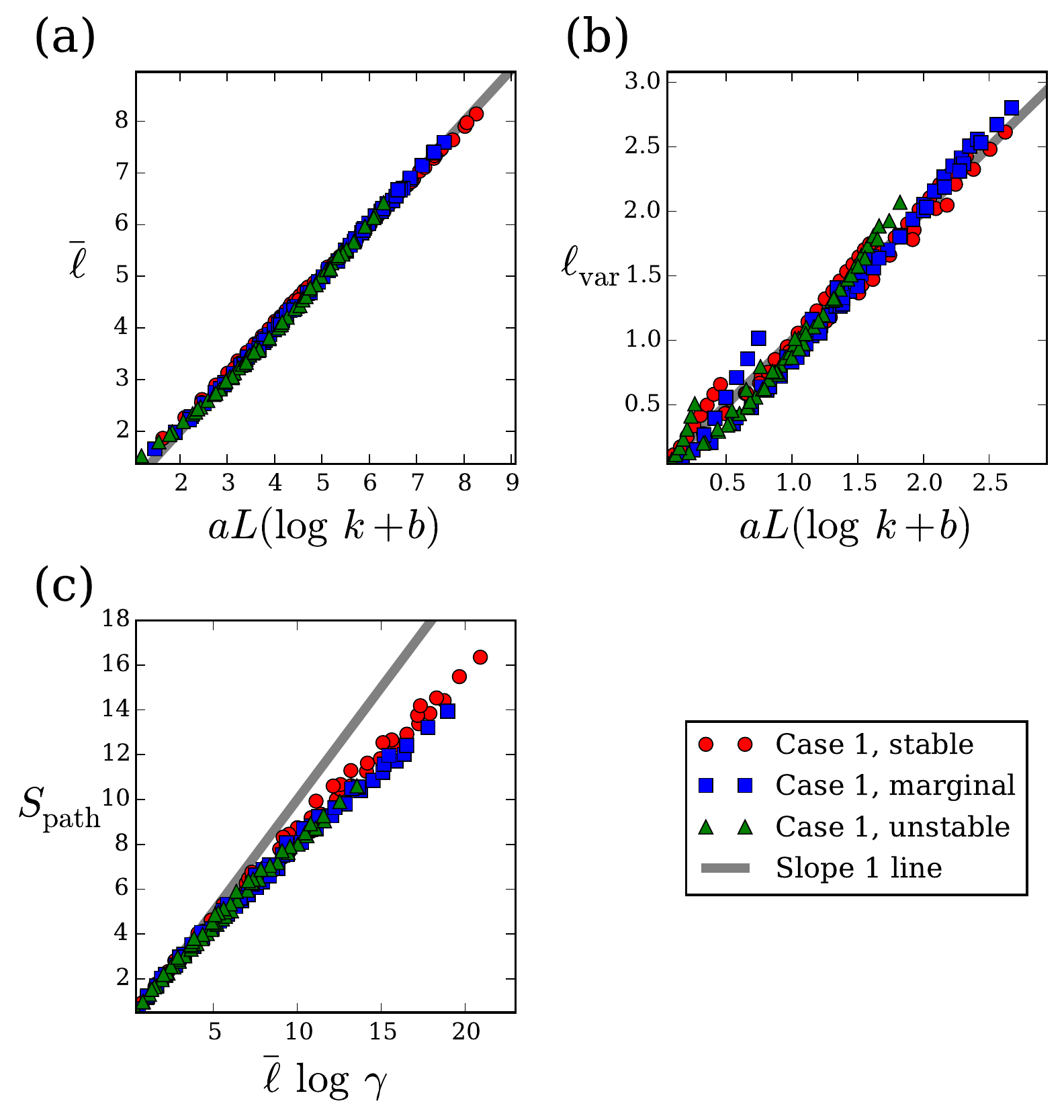}
\caption{
Scaling of path properties for three regimes of protein adaptation with direct selection for binding only (case 1). 
(a)~Mean path length (average number of substitutions) $\lbar$ and 
(b)~path length variance $\lvar$ versus $aL(\log k + b)$, 
where the parameters $a$ and $b$ are fitted separately for $\lbar$ and $\lvar$ and for stable, marginal, and unstable proteins.  
(c)~Path entropy $\Spath$ versus $\lbar \log\gamma$.  
All symbols are the same as in \fref{fig:case1_landscape_scaling}.
}
\label{fig:case1_path_scaling}
\end{figure}

     In \fref{fig:case1_path_scaling} we show the scaling of path statistics $\lbar$, $\lvar$, and $\Spath$.  We find that the strong-selection scaling relations describe these cases of the protein model very well, despite the complexities of the energy and fitness model relative to the simple Mount Fuji case.  The main discrepancy is in the path length variance, indicating that the distributions $\rho(\ell)$ are not as close to Poisson as in the Mount Fuji model.  We expect that this is mainly due to the small amount of epistasis present in the protein model.  Nevertheless, the scaling is accurate enough to extend the model to larger binding interfaces and a full amino acid alphabet.  For example, using the fitted coefficients $a$ and $b$ (\fref{fig:case1_path_scaling}a,b), we estimate $\lbar \approx 26$ and $\lvar \approx 9.6$ for a marginally-stable protein with $L_\phys = 12$ and $k_\phys = 20$.  Comparing these against the estimated $\lmin \approx 10$ (fitted as a linear function of $L(1-1/k)$; \fref{fig:case1_landscape_scaling}a), we see that many more substitutions than the minimum are likely.

\subsection{Cases 2 and 3: selection for folding stability}

\begin{figure}
\centering\includegraphics[width=\columnwidth]{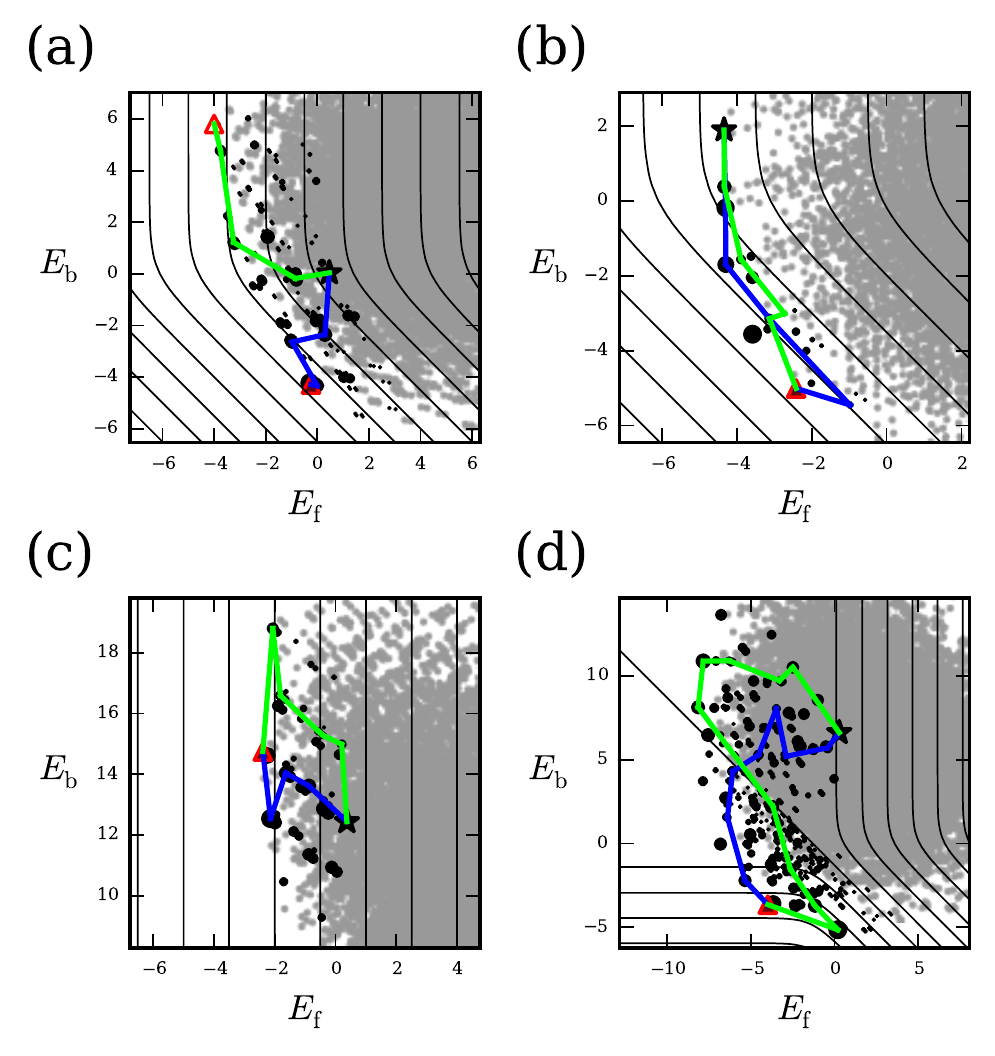}
\caption{
Example landscapes of protein adaptation with selection for folding stability (cases 2 and 3).  
(a)~Direct selection for folding only ($\fub = 1$, $\fuf = 0$) where both the old and new targets have potentially strong binding ($\EObone = \EObtwo = 3$ kcal/mol); 
(b)~same selection as (a) but where the old target has weak binding ($\EObone = 10$ kcal/mol) and the new target binds strongly ($\EObtwo = 3$ kcal/mol);  
(c)~same selection as (a) but where the old target has strong binding ($\EObone = 3$ kcal/mol) and the new target binds weakly ($\EObtwo = 10$ kcal/mol); and
(d)~direct selection for both binding and folding ($\fub = 0.9$, $\fuf = 0$) with marginal folding stability and binding strength ($\EOf = \EObone = \EObtwo = 5$ kcal/mol).
All symbols are the same as in \fref{fig:case1_landscapes}.  The coarse-grained sequence parameters are $L = 6$ and $k=5$, with \red{energy matrices rescaled using $L_\phys = 12$.}
}
\label{fig:case23_landscapes}
\end{figure}

     The fitness landscape changes qualitatively when there are additional selection pressures against misfolding beyond loss of function~\cite{Manhart2015}, e.g., for proteins that form toxic aggregates when misfolded~\cite{Bucciantini2002, Drummond2008, GeilerSamerotte2011}.  The first possibility is that the protein has a non-functional binding interaction ($\fub = 1$) but is deleterious when misfolded ($\fuf < 1$; ``case 2'').  Here the relative binding strengths of the old and new targets lead to different patterns of adaptation.  In \fref{fig:case23_landscapes}a, we show an example of adaptation when both the old and new targets have potentially strong (but non-functional) binding affinity, while \fref{fig:case23_landscapes}b shows an example when the old target has weak affinity while the new one has strong affinity.  \Fref{fig:case23_landscapes}c shows the case when the old target has strong affinity and the new target has little to no affinity.

     Finally, the most general case is to have distinct selection pressures on both binding and folding ($0 < \fub < 1$ and $\fuf < 1$; ``case 3''). Adaptation in this scenario often resembles binding-only selection (\fref{fig:case1_landscapes}), except when both binding and folding are of marginal strength (i.e., $\Ef \simeq 0$ and $\Eb \simeq 0$). In this case, the distribution of genotypes in energy space straddles a straight diagonal fitness contour, leading to a distinct pattern of evolutionary paths that gain extra folding stability first, only to lose it later as binding improves (\fref{fig:case23_landscapes}d).
     
\begin{figure}
\centering\includegraphics[width=\columnwidth]{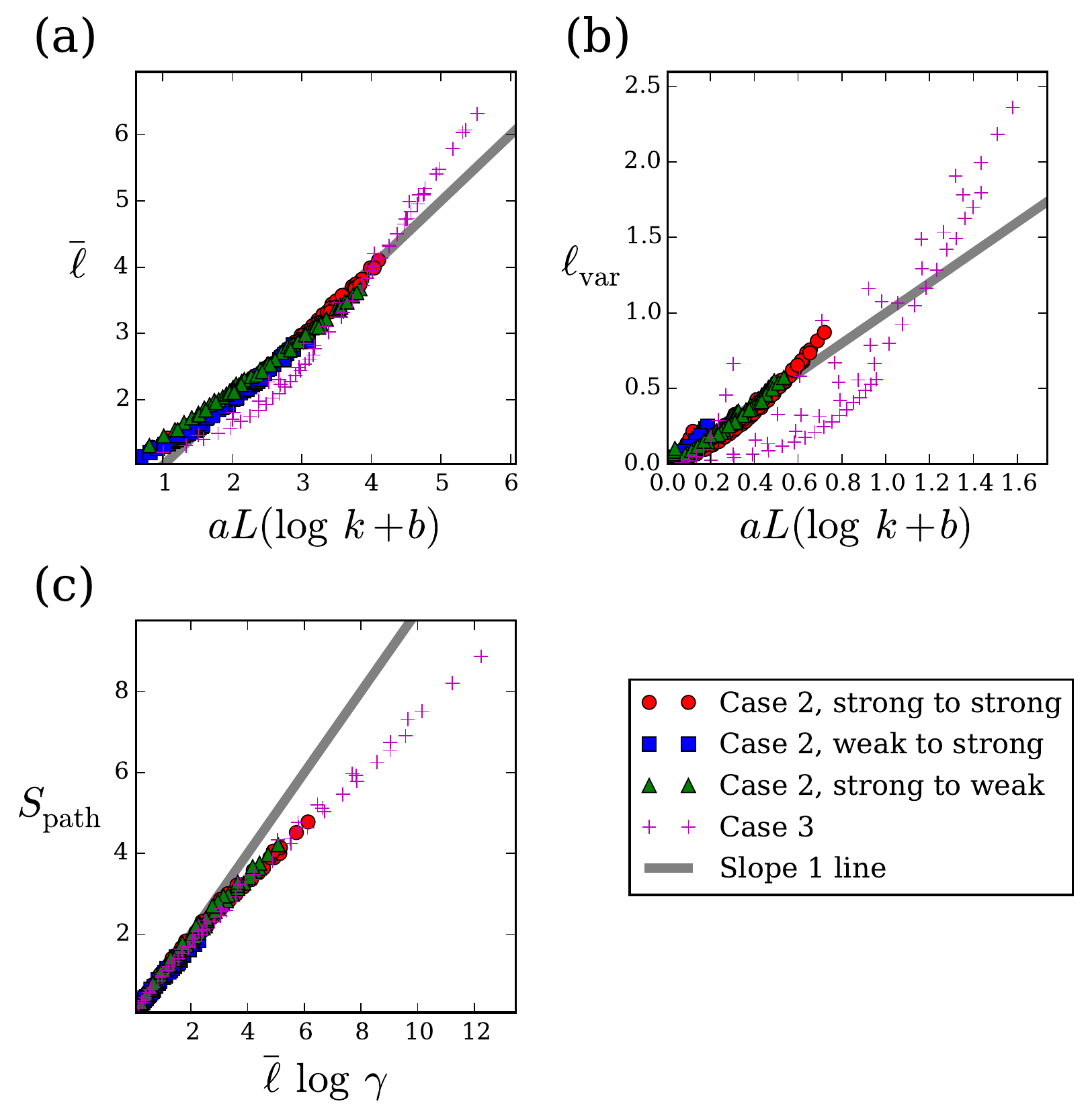}
\caption{
Scaling of path properties for protein adaptation with selection for folding stability (cases 2 and 3).  
Panels are the same as \fref{fig:case1_path_scaling} but with numerical data calculated using energy and fitness parameters matching examples in \fref{fig:case23_landscapes}: red circles are for case 2 (direct selection for folding only) proteins with strong binding to both old and new targets (\fref{fig:case23_landscapes}a); blue squares are for case 2 proteins with weak binding to the old target but strong binding to the new one (\fref{fig:case23_landscapes}b); green triangles are for case 2 proteins with strong binding to the old target but weak binding to the new one (\fref{fig:case23_landscapes}c); and purple crosses are for case 3 (direct selection for both binding and folding) proteins (\fref{fig:case23_landscapes}d).
}
\label{fig:case23_scaling}
\end{figure}

     We show the scaling properties of the evolutionary paths for cases 2 and 3 in \fref{fig:case23_scaling}.  In general, the predicted scaling relations are less accurate compared to binding-only selection (case 1, \fref{fig:case1_path_scaling}).  This is likely due to increased sign epistasis in these regimes.
Selection for both binding and folding (case 3) is particularly epistatic in the $\Ef \simeq 0$, $\Eb \simeq 0$ regime, leading to the largest deviations from the Mount Fuji scaling (\fref{fig:case23_scaling}).  On the other hand, the degree of epistasis here is still far from the maximally-epistatic, uncorrelated random landscape~\cite{Kauffman1987, Kingman1978}; in that model we should have $\lbar \sim \log L$~\cite{Flyvbjerg1992}, which is clearly not the case in our biophysical model.

\section{Discussion}

     Developing models of fitness landscapes based on the physics of proteins and other biomolecules has emerged as a powerful approach for understanding molecular evolution~\cite{Bloom2006, Lobkovsky2011, Schenk2013, Haldane2014, Serohijos2014b}.  However, the empirical nature of these models often makes explicit analytical treatments impossible, while the enormous size of sequence space often restricts numerical calculations or simulations to short sequences or reduced alphabet sizes.  While analyses with small $L$ and $k$ may preserve qualitative properties of the models, quantitatively extending these results to more realistic parameter values is essential for comparison with experimental data.  Here we have developed a scaling approach in which we empirically fit small $L$ and $k$ calculations to scaling relations to obtain precise quantitative properties of the model for arbitrarily large $L$ and $k$.  The scaling analysis confirms that small $L$ and $k$ calculations largely preserve qualitative properties of the model expected for realistic sequence spaces.  Although the scaling relations are derived for a much simpler, purely non-epistatic Mount Fuji model, they are surprisingly robust to the widespread magnitude epistasis and limited sign epistasis observed in the more realistic biophysical model of protein evolution.
     
     We also gain important conceptual insights from the scaling analysis.  In particular, we find that the neutral evolution scaling ($\lbar \sim \nseq = k^L$, $\lvar \sim \lbar^{~2} \sim k^{2L}$) holds even when selection is present, provided that it is not too strong ($Ns \leq 1$, \fref{fig:simple_model}a,b,c).  This means that the average number of substitutions to a global fitness maximum, even in the presence of weak selection, grows exponentially with $L$. 
On the other hand, strong selection enables populations to find the global maximum much faster: the mean path length scales with the logarithm of sequence space size, and the distribution of path lengths is approximately Poisson rather than exponential.  However, extremely strong selection ($Ns \approx 100$, \fref{fig:simple_model}e) is required for this more efficient behavior to take over.  Selection of this magnitude may be produced by sudden environmental changes, as in our model of protein adaptation~\cite{Manhart2015}.  When selection is of more moderate strength ($Ns \approx 10$), path length statistics are not simple functions of sequence space size (\fref{fig:simple_model}d).  We expect the more complex relations in this case to depend on the specific details of the landscape and evolutionary dynamics.

     Moreover, these insights are valuable for other types of random walks on complex landscapes, e.g., spin models where $L$ is the number of spins and $k$ is the number of individual spin states.  The scaling properties of first-passage paths have been well-studied for random walks in the absence of an energy or fitness landscape~\cite{Redner2001, Condamin2007}, but the effects of a landscape on scaling are less well-known.  Although the substitution dynamics of~\eref{eq:sub_rate} considered here are different from the typical dynamics used in spin models and other random walks (e.g., Metropolis Monte Carlo)~\cite{Yeomans1992}, we expect our qualitative findings to remain valid.  Thus we expect the pure random walk scaling ($T = \infty$) to hold for temperatures down to the size of the largest energy differences on the landscape.  There should be a non-trivial crossover regime at temperatures around the size of these landscape features, and then at small $T$ the $T = 0$ scaling will take over.  Investigating the nature of this crossover in both evolutionary and physical models is an important topic for future work.


\ack{AVM was supported by an Alfred P. Sloan Research Fellowship.}

\appendix

\section{Numerical algorithm for statistics of the path ensemble}
\label{sec:algorithm}

     We calculate statistical properties of evolutionary paths using an exact algorithm based on transfer matrices~\cite{Manhart2013, Manhart2014}.  
Let $Q(\s'|\s)$ be the jump probability defined by a rate matrix as in~\eref{eq:jump_prob}.  For each substitution $\ell$ and intermediate genotype $\s$, we calculate $P_\ell(\s)$, the total probability of all paths that end at $\s$ in $\ell$ substitutions, as well as $\Gamma_\ell(\s)$, the total entropy of such paths.  These quantities obey the following recursion relations:
     
\begin{eqnarray}
P_{\ell}(\s') &=& \sum_{\s} Q(\s'|\s) P_{\ell - 1} (\s), \nonumber\\
\Gamma_{\ell}(\s') &=& \sum_{\s} Q(\s'|\s) \left[ \Gamma_{\ell - 1} (\s) \right. \nonumber\\
				 & & \left. - (\log Q(\s'|\s)) P_{\ell - 1} (\s) \right],
\end{eqnarray}

\noindent where $P_0(\s) = 1$ if $\s$ is the initial state and $P_0(\s) = 0$ otherwise, and $\Gamma_0(\s) = 0$ for all $\s$.  Final states are treated as absorbing to ensure that only first-passage paths are counted.  We use these transfer-matrix objects to calculate the path ensemble quantities described in the text:

\beq
\rho(\ell) = \sum_{\s \in \mathcal{S}_\mathrm{final}} P_\ell(\s), \quad \Spath = \sum_{\ell=1}^\Lambda \sum_{\s \in \mathcal{S}_\mathrm{final}} \Gamma_\ell(\s),
\eeq

\noindent where $\mathcal{S}_\mathrm{final}$ is the set of final states.  The sums are calculated up to a path length cutoff $\Lambda$, which we choose such that $1 - \sum_{\ell = 1}^\Lambda \rho(\ell) < 10^{-6}$.  The time complexity of the algorithm scales as $\bigo(\gamma n\Lambda)$~\cite{Manhart2013}, where $\gamma$ is the average connectivity and $n$ is the total size of the state space.


\section{Mean path length in the strong-selection limit}
\label{sec:lbar_derivation}

     Since sites can be considered independent in the strong-selection limit, we need only calculate the mean path length for a single site with $k$ possible alleles.  A path begins at $\A_1$, and initially all $k$ alleles are of equal or higher fitness and are therefore accessible.  The first substitution can go to any $\A_j \in \{\A_2, \A_3, \ldots, \A_k\}$ with equal probability $(k-1)^{-1}$, after which there are $k-j+1$ remaining alleles.  Thus the mean path length $\lbar_k$ for $k$ alleles must satisfy the recursion relation
     
\beq
\lbar_k = 1 + \frac{1}{k-1} \sum_{j=2}^k \lbar_{k-j+1},
\label{eq:lbar_recursion}
\eeq

\noindent where $\lbar_1 = 0$.  This is satisfied by

\beq
\lbar_k = H_{k-1},
\label{eq:lbar_solution}
\eeq

\noindent where $H_n$ is the $n$th harmonic number defined by

\beq
H_n = \sum_{j=1}^n \frac{1}{j}.
\eeq

\noindent To prove this, we first note that

\begin{eqnarray}
\sum_{j=1}^n H_n & = n + \frac{n-1}{2} + \frac{n-2}{3} + \cdots + \frac{1}{n} \nonumber\\
& = \sum_{j=1}^n \frac{n+1-j}{j} \nonumber\\
& = (n+1) H_n - n \nonumber\\
& = (n+1) H_{n+1} - (n+1),
\label{eq:harmonic_sum}
\end{eqnarray}

\noindent where we have used the property $H_{n+1} = H_n + (n+1)^{-1}$.  Now we substitute $\lbar_j = H_{j-1}$ on the right-hand side of \eref{eq:lbar_recursion} and invoke \eref{eq:harmonic_sum} to obtain

\begin{eqnarray}
1 & + \frac{1}{k-1} \sum_{j=2}^k H_{k-j}  \nonumber\\
& = 1 + \frac{1}{k-1} \sum_{j=1}^{k-2} H_j \nonumber\\
& = 1 + \frac{1}{k-1} ((k-1)H_{k-1} - (k-1)) \nonumber\\
& = H_{k-1}.
\end{eqnarray}

\noindent This proves \eref{eq:lbar_solution} is the solution to the recursion relation.


\section{Distribution of path lengths in the strong-selection limit}
\label{sec:rho_derivation}

     Here we address the whole path length distribution $\rho(\ell)$ for a single site in the strong-selection limit.  With alleles ordered by fitness rank, a path of $\ell$ substitutions is of the form $\A_1 \to \A_{j_1} \to \cdots \to \A_{j_{\ell-1}} \to \A_k$, where $1 < j_1 < \cdots < j_{\ell-1} < k$.  Since all beneficial substitutions are equally likely in this limit, each jump probability out of allele $\A_j$ is $(k - j)^{-1}$.  Therefore the probability of taking a path of length $\ell$ is
     
\begin{eqnarray}
\rho(\ell) = & \frac{1}{k-1} \sum_{j_1=2}^{k-(\ell-1)} \frac{1}{k-j_1} \sum_{j_2=j_1+1}^{k-(\ell-2)} \frac{1}{k-j_2} \cdots \nonumber\\
& \sum_{j_{\ell-1}=j_{\ell-2}+1}^{k-1} \frac{1}{k-j_{\ell-1}}. 
\end{eqnarray}

\noindent The mean of this distribution is $\lbar = H_{k-1}$ as shown in \ref{sec:lbar_derivation}.  Here we obtain an approximate form for the whole distribution.  Define $\eps = k^{-1}$ and $x_i = j_i/k$.  For $k \gg 1$ ($\eps \ll 1$) we can take the continuum limit of the exact expression to obtain

\begin{eqnarray}
\rho(\ell) \approx & \frac{1}{k-1} \int_{2\eps}^{1 - (\ell-1)\eps} \frac{dx_1 }{1-x_1} \int_{x_1 + \eps}^{1 - (\ell-2)\eps} \frac{dx_2}{1-x_2} \cdots \nonumber\\
& \int_{x_{\ell-2} + \eps}^{1-\eps} \frac{dx_{\ell-1}}{1-x_{\ell-1}}.
\end{eqnarray}

\noindent By changing variables to $y_i = x_i - (i + 1)\eps$, we rewrite this as

\begin{eqnarray}
\rho(\ell) \approx & \frac{1}{k-1} \int_{0}^{1 - (\ell+1)\eps} \frac{dy_1}{1-y_1-2\eps} \nonumber\\
& \int_{y_1}^{1 - (\ell+1)\eps} \frac{dy_2}{1-y_2-3\eps} \cdots \nonumber\\
& \int_{y_{\ell-2}}^{1 - (\ell+1)\eps} \frac{dy_{\ell-1}}{1-y_{\ell-1}-\ell\eps}.
\end{eqnarray}

\noindent Each integral is dominated by its integrand's value near the upper limit.  However, because the domain of integration requires ordering of the $y_i$ variables ($0 < y_1 < y_2 < \dots < y_{\ell-1} < 1 - (\ell+1)\eps$), the integrand for $y_{\ell-1}$ has the greatest support near its upper limit.  Since the integrands are all similar near their lower limits, we thus approximate each integrand by the one for $y_{\ell-1}$:

\beq
\frac{1}{1 - y_i - (i+1)\eps} \approx \frac{1}{1 - y_i - \ell\eps}.
\eeq


\noindent This approximation allows us to use the identity

\begin{eqnarray}
\int_a^b dx_1~ f(x_1) \int_{x_1}^b dx_2~ f(x_2) \cdots \int_{x_{n-1}}^b dx_n~ f(x_n) = \nonumber\\
\quad \frac{1}{n!} \left(\int_a^b dx~ f(x) \right)^n.
\end{eqnarray}

\noindent Therefore,

\begin{eqnarray}
\rho(\ell) & \approx \frac{1}{k-1} \frac{1}{(\ell-1)!} \left( \int_0^{1 - (\ell+1)\eps} \frac{dy}{1 - y -\ell\eps} \right)^{\ell-1} \nonumber\\
& = \frac{\log^{\ell-1}(k-\ell)}{(\ell-1)! (k-1)}.
\end{eqnarray}

\noindent In the limit of $k \gg 1$ and $\ell/k \ll 1$, 

\begin{eqnarray}
\rho(\ell) & \approx \frac{(\log k + \log(1-\ell/k) )^{\ell-1}}{(\ell-1)! (k-1)} \\\nonumber
& \approx \frac{(\log k)^{\ell-1}}{(\ell-1)!} e^{-\log k}.
\end{eqnarray}

\noindent Thus $\rho(\ell)$ is approximately a Poisson distribution with mean and variance $\log k$.  This is consistent with the exact solution since $\lbar = H_{k-1} \approx \log k$ for large $k$.


\section{Size and connectivity of sequence space in the strong-selection limit}
\label{sec:size_connectivity}

     Each sequence $\s$ has $\sum_{j=1}^k (k-j) n_j(\s)$ possible beneficial mutations in the Mount Fuji model~\eref{eq:MF_fitness}.  Thus the connectivity averaged over all sequences is
     
\begin{eqnarray}
\gamma & = \frac{1}{k^L} \sum_{n_1, n_2, \ldots, n_k} {L \choose n_1, n_2, \ldots, n_k} \sum_{j=1}^k (k-j)n_j \nonumber\\
& = \sum_{j=1}^k (k-j) \frac{L}{k} \nonumber\\
& = \frac{1}{2} L(k-1). 
\end{eqnarray}

\noindent We can also determine the average connectivity of the accessible sequences starting from a random initial sequence.  We first consider a single site.  The initial allele $\A_j$ is chosen with probability $1/k$, leaving $k - j +1$ accessible alleles.  Thus the average connectivity of this accessible space is 

\begin{eqnarray}
\gamma & = \frac{1}{k} \sum_{j=1}^k \sum_{i = j}^k \frac{1}{k-j+1} (k - i) \nonumber\\
& = \frac{1}{4} (k-1).
\end{eqnarray}

\noindent Since multiple sites contribute additively to the connectivity, the total average connectivity of the accessible space is $L(k-1)/4$.

     Starting from the sequence with minimum fitness, all $k^L$ sequences are accessible in the strong-selection limit.  More generally, if the population begins at sequence $\s$, there are $\prod_{j=1}^k (k-j+1)^{n_j(\s)}$ accessible sequences, including $\s$ itself. If the initial sequence is chosen at random, then the average number of accessible sequences is
     
\begin{eqnarray}
n_\mathrm{seq} & = \frac{1}{k^L} \sum_{n_1, n_2, \ldots, n_k} {L \choose n_1, n_2, \ldots, n_k} \prod_{j=1}^k (k-j+1)^{n_j(\s)} \nonumber\\
& = \sum_{n_1, n_2, \ldots, n_k} {L \choose n_1, n_2, \ldots, n_k} \prod_{j=1}^k \left(1 - \frac{(j-1)}{k} \right)^{n_j(\s)} \nonumber\\
& = \left( \sum_{j=1}^k \left(1 - \frac{(j-1)}{k} \right) \right)^L \nonumber\\
& = \left( \frac{k+1}{2} \right)^L.
\end{eqnarray}


\section*{References}
\bibliographystyle{unsrt}
\bibliography{Bibliography}

\end{document}